\documentclass[10pt, conference]{IEEEtran}
\usepackage{cite} \usepackage{amsmath,amssymb,amsfonts} \usepackage{algorithmic}
\usepackage{graphicx} \usepackage{textcomp} \usepackage{xcolor} \def\BibTeX{{\rm
B\kern-.05em{\sc i\kern-.025em b}\kern-.08em
T\kern-.1667em\lower.7ex\hbox{E}\kern-.125emX}} 

\begin{document}

\title{Supervised Machine Learning Techniques for Trojan Detection with Ring
Oscillator Network\\ }

\author{\IEEEauthorblockN{Kyle Worley and Md Tauhidur Rahman}
\IEEEauthorblockA{\textit{Department of Electrical and Computer Engineering} \\
\textit{The University of Alabama in Huntsville, Huntsville, AL}\\ (kaw0035,
tauhidur.rahman)@uah.edu} 
 }

\maketitle

\begin{abstract} 

With the globalization of the semiconductor manufacturing process, electronic
    devices are powerless against malicious modification of hardware in the
    supply chain. The ever-increasing threat of hardware Trojan attacks against
    integrated circuits has spurred a need for accurate and efficient detection
    methods. Ring oscillator network (RON) is used to detect the Trojan by
    capturing the difference in power consumption; the power consumption of a
    Trojan-free circuit is different from the Trojan-inserted circuit. However,
    the process variation and measurement noise are the major obstacles to
    detect hardware Trojan with high accuracy. In this paper, we quantitatively
    compare four supervised machine learning algorithms and classifier
    optimization strategies for maximizing accuracy and minimizing the false
    positive rate (FPR). These supervised learning techniques show an improved
    false positive rate compared to principal component analysis (PCA) and
    convex hull classification by nearly 40\% while maintaining $>$ 90\% binary
    classification accuracy.  

\end{abstract}

\begin{IEEEkeywords} hardware Trojans, machine learning, supervised learning,
ring oscillator network \end{IEEEkeywords}

\section{Introduction} While the transition from vertically integrated supply
chains to horizontally integrated has decreased costs for integrated circuit
(IC) designers; the "fabless" approach comes with the steep price of trust
\cite{Trojan:Tehrani, Trojan:Bhunia, Trojan:JV, Trojan:hassan, TrojanDet:Karri,
CSST, Supplychain:sec:rahman, CSST:NATW, Hardware:IP}. Semiconductor designers
now must trust their intellectual property (IP) to multiple parties in order to
have their ICs manufactured at foundries \cite{Hardware:IP, CSST}. Not only do
they run the risk of having their IP stolen, but it is not uncommon for
untrusted system integrators and foundries to insert hardware Trojans before
shipping the final product \cite{Trojan:Tehrani, Trojan:Bhunia, Trojan:JV,
Trojan:hassan}. These Trojans are capable of leaking sensitive information,
disabling key portions of the IC, self-destructing the chip, or hindering
performance \cite{Trojan:Tehrani, Trojan:Bhunia}. This has driven the need for
fast, accurate, and simple methods of detecting infected ICs before they are
able to taint the supply chain.

With this rising need, some research in the field of hardware security  has been
focused on finding optimal methods of detecting and classifying Trojans.
Initial research suggested the use of semi-invasive strategies such as scanning
electron microscopy (SEM) for failure analysis. However, this is expensive and
time consuming for it to be applied to every IC.  Using netlist failure
detection techniques was also unsuccessful due to Trojans that add functional
logic remaining undetected.  

The most promising technique relies on the use of side channel information as it
is non-invasive and can be done quickly.  By monitoring side channel information
from an IC power grid it is possible to detect Trojans due to their additional
activity \cite{b3}.  In \cite{b1}, the authors developed a ring oscillator
network (RON) in a chip's power grid for hardware Trojan detection. The
increased switching activity from Trojan activation will manifest itself in
decreased RO frequencies due to the variable voltage drop in the chip's power
network. Using Principal Component Analysis (PCA) and convex hull classification
(\cite{PCA,convexHull}) they were able to achieve greater than 80\%
classification accuracy with a false positive rate of 50\%.

This was improved upon in \cite{b4} using a genetic algorithm (GA) for feature
reduction and a support vector machine (SVM) for classification.  Feature
reduction allows machine learning algorithms to reduce the feature space and
decrease training time and the possibility of over fitting.  The genetic
algorithm is built upon the idea of natural selection where the best features
will "survive" through each generation. When the end of the algorithm is reached
you will be left with the optimal feature set.  This in addition to the use of
SVM resulted in 99.6\% classification accuracy and a reduced false positive
rate. However, \cite{b4} still suffers from a large FPR.  

In this paper, we present a supervised machine learning approach for the
classification of Trojan free and infected ICs using a RON. The results show
that we maintain similar accuracy to previous work in addition to reducing the
FPR by using the K-Nearest Neighbors (KNN), Support Vector
Machine (SVM), Naive Bayes, and ensemble classification algorithms \cite{KNN,
SVM,NB, EL}. Experimental results show detection accuracy $>$ 88\% with some
classifiers even reaching 97.4\%.  Low false positive rates (FPR) were also
achieved and in the case of two classifiers a $\sim$ 0\% FPR was reached.

The rest of this paper will be laid out as follows: Section \ref{sec:background}
will provide all necessary background information, Section III will discuss my
proposed method of classification, and Section IV will contain the results and
discussion of our proposed method. Section \ref{sec:conclusion} will conclude
our proposed work. 

\section{Background and Experimental Set-up} \label{sec:background}
\subsection{Hardware Trojan and Related Work} Hardware Trojans are malicious
modifications made by attackers during the design and manufacturing process
\cite{Trojan:Tehrani, Trojan:Bhunia, Trojan:hassan, TrojanDet:Karri}. Trojans
can be used to degrade performance, steal information, or block functionality of
an IC \cite{Trojan:Tehrani, Trojan:Bhunia}. These unwanted circuit additions are
often hard to detect because they are not always triggered or activated by
standard test procedures \cite{Trojan:Tehrani, Trojan:Bhunia}. Trojans also come
in a wide variety of formats so no single filter can catch them all. These
Trojans are unwanted and pose a risk to the chip owners of receiving secretly
modified hardware that could lead to devastating consequences
\cite{Trojan:Tehrani, Trojan:Bhunia}. A hardware Trojan can be classified into
three main categories according to their physical, activation, and action
characteristics \cite{Trojan:Tehrani, Trojan:Bhunia}. The physical
characteristics category describes the various hardware manifestations of
Trojans according to their shape and size; the activation characteristics
describe the conditions, which activate the Trojans, and action characteristics
refer to the behavior of the Trojans \cite{Trojan:Tehrani, Trojan:Bhunia}.

Current Trojan detection methods largely focus on (i) functional testing and
(ii) side channel analysis \cite{TrojanDet:Karri, TrojanDet:Fareena, b1, b2, b3,
b4}. Functional verification is the attempt to activate Trojans by applying test
vectors and comparing the responses with the correct results
\cite{Trojan:Tehrani,b3}.  The difficulty with this approach is the rarity of
which some hardware Trojans are activated. It is nearly impossible to explore
every possible state of a circuit and search for Trojan activity
\cite{TrojanDet:Karri}. Whereas, side-channel analyses detect the HT by
analyzing the physical characteristics of the IC chip such as transient current,
leakage current, delay, energy, heat generation, or EM radiation
\cite{TrojanDet:Karri, TrojanDet:Fareena, TrojanDet:Zhou, b1, b2, b3}.  In both
approaches, the outputs of circuit under test are compared with the outputs of a
golden circuit. Typically, the adversary would design a Trojan to evade
detection by ensuring rare activation to evade logic testing and
minimal physical characteristics, like size, to escape side channel based
testing. Backside optical imaging of the fabricated chip enables extraction of
the full standard cell layout of the chip with the watermarks, which in turn can
be validated with image processing against the expected simulated layout to
detect any changes made to accommodate hardware Trojans \cite{TrojanDet:Zhou}. A
challenge in backside imaging is obtaining a high enough spatial resolution for
an accurate representation of a nanometer-scale circuit \cite{TrojanDet:Zhou}.

\subsection{Supervised Learning} In the field of machine learning there are two
main approaches in use: supervised and unsupervised learning \cite{ML}.
Unsupervised learning is outside the scope of this paper and will not be
discussed further for the sake of brevity. Supervised learning algorithms work
under the assumption the training data is labeled before being processed by the
algorithm.  By labeling the data, the algorithm then knows the desired output
for the given input set and can create a hypothesis for determining the desired
output for future inputs \cite{ML}.  Within the topic of supervised learning
exist two problem types: regression and classification. Regression
algorithms are going to map input values to a real output value, e.g., 
predicting stock market prices given a feature set.  On the other hand,
classification algorithms will place a set of input data points into one or more
"classes", e.g. Trojan free or Trojan infected.

In this paper, a binary classification (\cite{binaryclass, ML}) approach is used
to classify each IC as either Trojan free or Trojan infected. In order to
properly train the classifier we must operate under the assumption we have data
from both Trojan free and Trojan infected circuits. Obtaining known Trojan free
ICs is a challenge in and of itself, but knowing which ICs are infected with
Trojans will require some other method of detection until enough data can be
collected to train a classifier.

\subsection{K-Nearest Neighbors} 
\begin{figure}[htb] \centerline{\includegraphics[scale=0.25]{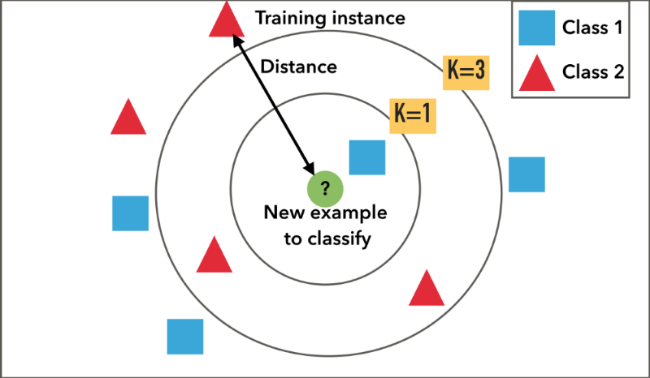}}
    \caption{An example of the KNN algorithm showing the effect the value of $k$
    has on classification. Notice that the new example will be classified as
    class 1 if $k$ = 1, but class 2 if $k$ = 3 \protect\cite{b5}.} \label{knn}
\end{figure}

One of the simplest and yet most popular machine
learning algorithms is k-nearest neighbors (KNN).  When used for classification
the $k$ nearest training samples in the feature space are used to classify the new
point through a simple majority vote.  This simplicity does come with the cost
of longer classification times for larger data sets. The value $k$ is usually defined as a
positive integer, and in the case of binary classification it is useful to set k
as an odd number to prevent a split decision.  The distance metric can be any
method of calculating distance, but Euclidean distance is often used.  It can be
defined as follows: \begin{equation} d(x,y) = \sqrt{(x_1 - y_1)^2 + \dots + (x_n
- y_n)^2} \end{equation} As can be seen in \ref{knn}, the value of $k$ will have
an effect on the classifiers performance.  By comparing the mean error and
accuracy values for a series of $k$ values it is possible to find the most optimal
$k$ value for a data set.  

In this paper, the classifier was trained using a range of $k$ values from 1 to 40
and the value with the best FPR and accuracy was selected without being
over fitted.  It is usually safe to select a value of $k$ near the square root of
the number of training samples.  However, low values of $k$ will lead to a
classifier that performs worse with noisy data, and high values of $k$ can lead to
over fitting the classifier to the training data.

\subsection{Support Vector Machine} Another popular machine learning algorithm
is known as the support vector machine (SVM) \cite{SVM}.  While not as simple as
KNN it is much more powerful for classification and regression applications.
The training of the SVM consists of finding the optimal hyperplane that will
linearly classify data points with the largest margin possible between the two
classes of data points.  However, not all data can be linearly separated by a
hyperplane in which case we must apply a "kernel trick" to transform the feature
space.  

\begin{figure}[htb] \centerline{\includegraphics[scale=0.5]{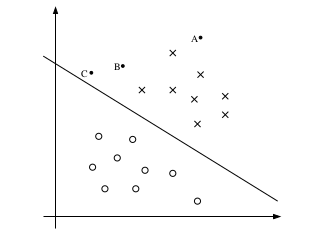}}
    \caption{Example of a SVM hyperplane separating two classes of data. In this
    example A, B, and C would be classified by computing the dot product.  They
    all wold be classified as class 'x' \protect\cite{b6}} \label{svm}
\end{figure}

If we define our training data as a set of points in the form of
$(\vec{x_1},y_1),\dots,(\vec{x_n},y_n)$, where $\vec{x_i}$ is a vector and $y_i$
is -1 or 1 to represent the class of $\vec{x_i}$, then we can define our
hyperplane as satisfying the following equation \cite{SVM}: \begin{equation}
\vec{w}\cdot\vec{x}-b = 0 \end{equation} If the data set is linearly separable
then one class can be defined as anything on or above the boundary
$\vec{w}\cdot\vec{x}-b=1$ and the other class can be defined as anything on or
below $\vec{w}\cdot\vec{x}-b=-1$.  Now in order to train the SVM we want to
minimize the difference between these two hyperplanes so that the margin between
the two classes is maximized.  Thus the problem simplifies down into:
\begin{equation} \min\limits_{n^1 \dots n^i} \text{ for }
y_i(\vec{w}\cdot\vec{x_i}-b) \geq 1 \end{equation} The classifier will then be
defined by $\vec{w}$ and $b$.  An example of a hyperplane used to separate two
classes of data can be seen in \ref{svm}.

This method will only work for linearly separable data and the classification of
other data requires the replacement of the dot product with a nonlinear kernel
function, thus the name "kernel trick".  By using the kernel function we can now
put a hyperplane in our higher dimensional nonlinear feature space.  In this
paper a Gaussian radial base function was used.  \begin{equation}
k(\vec{w_i},\vec{x_j}) = exp(-\gamma\Vert\vec{x_i}-\vec{x_j}\Vert^2) \text{ for
} \gamma > 0 \end{equation}

\subsection{Naive Bayes} The construction of classifiers using Naive Bayes is a
relatively simple process that can produce highly accurate and fast
classification results using a probabilistic approach\cite{NB}. Using Bayes
theorem we can generate the probability that a data point will belong to a class
$C_k$ given the presence of one of the features of the data \cite{NB}.

\begin{equation} P(C_k|x) = \frac{P(x|C_k)P(C_k)}{P(x)} \end{equation}

However, when trying to build a classifier we are interested in the probability
a data point belongs in class given multiple features.  Using the chain rule
Bayes theorem can be expanded to account for this.  Assuming the $n$ features in
the data set can be represented as $X = (x_1, x_2, ... , x_n)$ then it follows
\cite{NB}:

\begin{equation} P(C_k|x_1,...,x_n) =
\frac{P(x_1|C_k)P(x_2|C_k)...P(x_n|C_k)P(C_k)}{P(x_1)P(x_2)...P(x_n)}
\end{equation}

Now if we assume the conditional independence of the features in the set:

\begin{equation} P(C_k|x_1,...,x_n) = \frac{P(C = C_k)\prod_{i} P(x_i|C =
C_k)}{\sum_{j}P(C = C_j)\prod_{i}P(X_i|C = C_j} \end{equation}

Finally, to create a classification rule we must have a way of making decisions.
Using a \textit{maximum a posteriori} rule, or simply stated choosing the most
probable outcome, we can decide how to assign class labels to data points.

\begin{equation} y = argmax P(C_k)\prod_{i=1}^{n}P(x_i|C_k) \end{equation}

Naive Bayes can be applied in one of three ways to estimate the likelihood of
the features.  A Gaussian classifier will assume the features are distributed on
a Gaussian distribution, Multinomial will assume multinomially distributed data,
and Bernoulli will assume binary-valued features.  The Gaussian classifier was
selected in this paper due to the continuous nature of the data set.

\subsection{Ensemble Learning} Ensemble learning is another technique for
producing better prediction results using machine learning algorithms \cite{EL}.
It operates under the assumption that by combining the predictive power of single
algorithms it is possible to increase the overall possible predictive power.
Several popular strategies include voting, bagging, stacking, boosting, and
"bucket of models".  In this paper, we will only be implementing a simple voting
method.  This is done by taking the output of each of the three classifiers and
using it as a vote. The class with the most votes will then be the output of the
classifier.  Theoretically, given an odd number of classifiers in the ensemble a
decision should be made every time that represents the best of each classifier
\cite{EL}.  However, if the ensemble is made up of an even number of classifiers
situations can arise resulting in split decisions.  This is said to be an
unstable decision.  This can be mitigated through the use of either an odd
number of classifiers or using weighted voting to reduce the possibility of a
split decision.  

\subsection{Ring Oscillator Network and Trojan Detection} 
\begin{figure}[htb] \centerline{\includegraphics[scale=0.58]{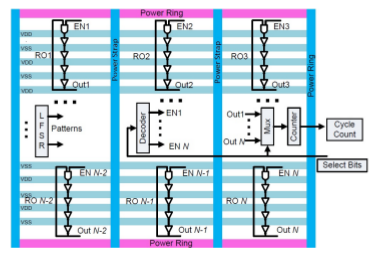}}
    \caption{The ring oscillator network used for Trojan detection.  While 8 ROs
    are used in this configuration the structure will differ based on the power
    network of the IC you are trying to protect \protect\cite{b1,b4}}
\label{ron} \end{figure}

Recent work has shown
that a ring oscillator (RO) network (RON) connected to the power supply
structure of an IC can be used to detect hardware Trojan activity.  As shown in
\ref{ron}, ROs consisting of inverters and a NAND gate for activation control
are placed in a vertical orientation within the power structure of an IC.  The
ROs are then provided test patterns from a linear feedback shift register and a
decoder.  These outputs are then selected using a multiplexer and a counter
registering the number of oscillations from the selected RO. The RO's frequency
can then be derived from the number of oscillations.  Any Trojan inserted into
an IC will result in extra noise in the power supply structure that would not
otherwise be present in a "golden" chip.  By injecting the same test patterns
into every IC the Trojans should at least partially active and thus cause extra
noise.  Since a RO's frequency is directly related to its power supply voltage
this Trojan caused power supply noise should propagate to the RO's frequency and
result in differing measurements between clean and infected ICs \cite{b1,b4}.  

However, the frequency differences are not always discernible to the human eye
nor to simple algorithmic classification strategies due to process variations
and other factors.  In \cite{b1} Principal Component Analysis (PCA) was used as
a means of feature reduction.  The data set contained the frequency data from 8
ROs, but through feature reduction could be accurately represented with just 3.
A simple convex hull classification method was then used to classify each IC as
either Trojan free or into one of the 23 Trojan categories. While the RON is
successful at detecting the difference between Trojan free and Trojan infected
circuits the false positive rate was nearly 50\%.  Using the data collected from
the RON we will try to improve on this false positive rate while maintaining
above 90\% classification accuracy.

\subsection{Experimental Set-up}\label{sec:experimental} We conducted our
experiments on eight FPGA boards (Nexys4 DDR development board \cite{FPGA}).
Each FPGA board is divided into four separate regions to increase the sample
size. Each region is considered as an individual IC and Trojan, and the RON
architecture is implemented in only a single portion at a time in order to make
sure that one portion (or an individual IC) does not interfere another. We used
a total of eight 41-stage ROs in each portion (i.e., IC). We distributed
combinational and sequential Trojans (\cite{TRUSTHUB}) in one portion randomly.
We used several Trojan benchmarks from Trusthub \cite{TRUSTHUB}. We measured the
average RO frequency at room temperature and nominal operating voltage from 50
measurements (with Trojan and without Trojan) to cancel out the measurement
noise. We included ITC-99 (\cite{ITC99}) benchmarks for normal operation.    

\section{Method} \label{sec:method} The method we will use in this paper is to
use the four previously discussed supervised classification approaches and
optimize them for accuracy and a low FPR. The main motivation for this is to
reduce potential waste of Trojan free ICs that would otherwise be discarded due
to being classified as infected.  However, accuracy must be maintained to
prevent Trojans from being introduced into the supply chain.  

In order to do this, from the collected data, each chip has readings for two
"golden" or Trojan free samples and 23 Trojan inserted samples.  The data was
collected using the test setup described in \ref{sec:experimental}.  This data
was then be labeled accordingly and used to train the classifiers.

The KNN classifier will then be optimized by finding the best $k$ value for
maintaining accuracy and minimizing the FPR.  By training the KNN classifier on
a range of $k$ values and different training sample sizes we were able to select
the best value for our data set.  The SVM classifier will be optimized using
two slack values pertaining to the Gaussian kernel function, $C$ and $\gamma$.
$C$
can be considered the weight correct classification has over maximizing the
margin between the two classes.  Gamma is the inverse of the variance of our
Gaussian function. Thus, a small $\gamma$ will lead to a large variance and
points could be similar even if they are not close together and vice versa.  In
order to find the optimal $\gamma$ and $C$ values we have used a grid search
method in which a given set of values is exhaustively run through until the best
values for the data set are found.  The Naive Bayes Gaussian classifier will not
be tuned using any parameters.  Each of the three classifiers will then be
combined in a simple voting ensemble in the following combinations:
KNN+SVM+Naive Bayes, KNN+SVM, KNN+Naive Bayes, and SVM+Naive Bayes.  The KNN and
SVM classifiers will retain the same optimization parameters as they had being
trained individually.

\begin{figure}[htb]
    \centerline{\includegraphics[scale=0.6]{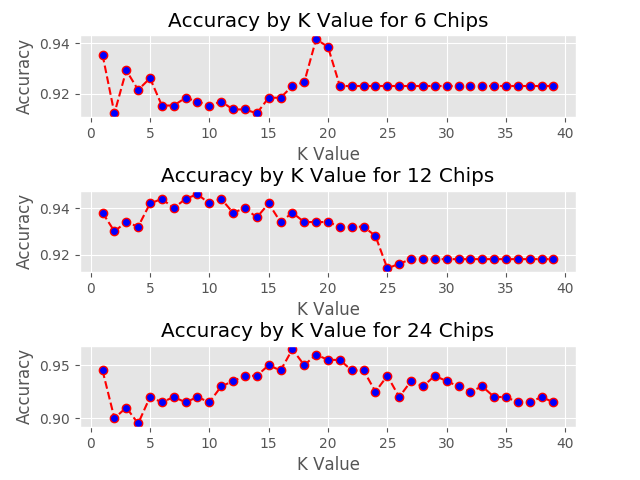}}
    \caption{A plot showing the effect the value of $k$ has on the classifier
    accuracy.  As can be seen a $k$ value of 2 provides enough accuracy without
    being over fitted.} \label{accuracyByK} \end{figure}

\begin{figure}[htb]
    \centerline{\includegraphics[scale=0.6]{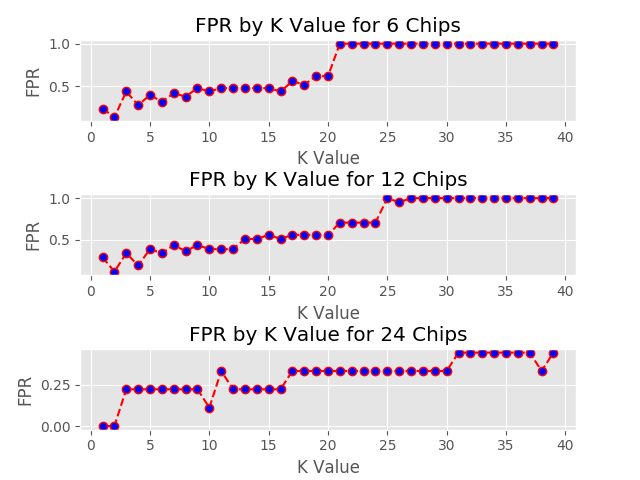}} \caption{A
    plot showing the effect the value of $k$ has on the classifier false positive
    rate. As shown higher values of k will lead to over fitting and higher false
positive rates.} \label{fprByK} \end{figure}

\section{Results} Following the method above each classifier was trained and
optimized for three different sized data sets consisting of 6 chips, 12 chips,
and 24 chips.  Each sample size was then repeated for 20 trials and the average
accuracy, false positive rate (FPR), false negative rate (FNR), true negative
rate (TNR), and true positive rate (TPR) were calculated and recorded as
follows: \begin{equation} TPR = \frac{TP}{TP + FN} \end{equation}
    \begin{equation} TNR = \frac{TN}{TN + FP} \end{equation} \begin{equation}
        FNR = \frac{FN}{FN + TP} \end{equation} \begin{equation} FPR =
        \frac{FP}{FP + TN} \end{equation}

The optimization step of the training led to the discovery of useful properties
of our data set.  Initial estimates for the value of $k$ when training the KNN
classifier used the square root of the number of samples in the training data
set.  While this resulted in a very accurate classifier it came at the cost of a
FPR greater than or equal to 50\%.  This is most likely a result of the data set
having little noise and being very prone to over fitting.  In Figure
\ref{accuracyByK}, the accuracy for a range of $k$ values is depicted, note that as
the value increases the accuracy tends to plateau as a result of over-fitting.
Figure \ref{fprByK} shows the same plateau for the FPR.  Since we want to avoid
over-fitting and lower $k$ values perform well we can assume the data set is not
noisy.  This led to the decision to use a $k$ value of 2 for every sample size.  This maintained the
greater than 90\% accuracy benchmark and had a best case FPR of only 9.4\%, a
near 40\% decrease compared to PCA and convex hull classification. Even with
small training sets the KNN maintained a FPR under 20\% (Table
\ref{knnResults}).  

\begin{table}[htbp] \caption{KNN Classifier Results} \begin{center}
    \begin{tabular}{|c|c|c|c|} \hline
        \textbf{Metric}&\multicolumn{3}{|c|}{\textbf{Sample Size}} \\
        \cline{2-4} & \textbf{6 Samples}& \textbf{12 Samples}& \textbf{24
        Samples} \\ \hline TNR & 0.813 & 0.815 & 0.906 \\ \hline FPR & 0.187 &
    0.185 & 0.094 \\ \hline FNR & 0.075 & 0.063 & 0.051 \\ \hline TPR & 0.916 &
    0.927 & 0.745 \\ \hline Accuracy & 0.916 & 0.927 & 0.945 \\ \hline
    \end{tabular} \label{knnResults} \end{center} \end{table}

Optimizing the SVM proved to be more difficult than the KNN classifier.  The
    grid search was quick to converge on a $C$ value of 1 and $\gamma$ value of 0.1,
    but the FPR left much to be desired.  As can be seen in Table
    \ref{svmResults}, the SVM is very accurate but when trained on fewer samples
    it struggles with a high FPR. Using a balancing optimization it was still
    able to achieve a 97.4\% classification accuracy and a 7.1\% FPR (Table
    \ref{svmResults}) and outperform convex hull and approach the results
    achieved in \cite{b4}.  This leads me to believe that with a larger data set
    and increased training set sizes the SVM could become more accurate and
    reduce the FPR even further.  Unfortunately, it is not always possible to
    have large data sets due to factors outlined above.

\begin{table}[htbp] \caption{SVM Classifier Results} \begin{center}
    \begin{tabular}{|c|c|c|c|} \hline
        \textbf{Metric}&\multicolumn{3}{|c|}{\textbf{Sample Size}} \\
        \cline{2-4} & \textbf{6 Samples}& \textbf{12 Samples}& \textbf{24
        Samples} \\ \hline TNR & 0.445 & 0.605 & 0.929 \\ \hline FPR & 0.555 &
    0.355 & 0.071 \\ \hline FNR & 0.017 & 0.023 & 0.023 \\ \hline TPR & 0.983 &
    0.977 & 0.977 \\ \hline Accuracy & 0.940 & 0.946 & 0.974 \\ \hline
    \end{tabular} \label{svmResults} \end{center} \end{table}

Despite the many operating assumptions the Naive Bayes classifier is a
    very powerful but simple and fast method.  With no optimization the
    classifier produced results that were slightly less accurate compared to the
    other classifiers.  At the 6 chip sample size the classifier was 88.3\%
    accurate but had only a 6.9\% FPR. The accuracy only dropped 0.1\% when
    increasing the training sample size to 12 chips, but the FPR dropped to
    6.1\%, the lowest FPR of any non-ensemble classifier (Table
    \ref{nbResults}).  The Naive Bayes classifier produced the best results in
    term of FPR but was held back by a higher FNR which led to reduced accuracy.
    In theory, this could be reduced by tuning the decision threshold, but would
    most likely result in the FPR increasing.  

\begin{table}[htbp] \caption{Naive Bayes Gaussian Classifier Results}
    \begin{center} \begin{tabular}{|c|c|c|c|} \hline
        \textbf{Metric}&\multicolumn{3}{|c|}{\textbf{Sample Size}} \\
        \cline{2-4} & \textbf{6 Samples} & \textbf{12 Samples}& \textbf{24
        Samples} \\ \hline TNR & 0.931 & 0.955 & 0.939 \\ \hline FPR & 0.069 &
        0.045 & 0.061 \\ \hline FNR & 0.121 & 0.124 & 0.127 \\ \hline TPR &
        0.879 & 0.876 & 0.873 \\ \hline Accuracy & 0.883 & 0.882 & 0.873 \\
    \hline \end{tabular} \label{nbResults} \end{center} \end{table}

When using ensemble learning the hope is the results are better than that of
    each of the individual classifiers by themselves.  However, it also runs the
    risk of the opposite occurring.
    We encountered both situations while training the ensembles. The ensemble
    containing all three classifiers performed better than the lone SVM
    classifier at the lower training sample sizes.  Yet, it was outperformed at the
    24 chip sample size (Table \ref{allResults}).  The Naive Bayes and KNN/SVM
    ensembles had the lowest overall FPRs of all classifiers, but struggled to
    beat the desired 90\% binary classification accuracy threshold (Tables
    \ref{snResults} \& \ref{knResults}).  This can be attributed to the Naive
    Bayes classifier's characteristics dominating those of the other
    classifiers.  Despite the lower accuracy, at the 24 chip training sample
    size both ensembles had a 0\% FPR.  Overall, the best ensemble method was
    the combination of the SVM and KNN classifiers (Table \ref{skResults}).  At
    the lower training sample sizes the FPR was only 19.6\% and 16.4\%, but kept
    an accuracy of 92.1\% and 93.0\% respectively.  At the 24 chip training
    sample size the FPR was 0.03\% higher than the SVM alone but with a 3.4\%
    accuracy loss.  

\begin{table}[htbp] \caption{SVM+KNN+NB Ensemble Classifier Results}
    \begin{center} \begin{tabular}{|c|c|c|c|} \hline
        \textbf{Metric}&\multicolumn{3}{|c|}{\textbf{Sample Size}} \\
        \cline{2-4} & \textbf{6 Samples} & \textbf{12 Samples}& \textbf{24
        Samples} \\ \hline TNR & 0.785 & 0.796 & 0.908 \\ \hline FPR & 0.215 &
        0.204 & 0.092 \\ \hline FNR & 0.066 & 0.062 & 0.055 \\ \hline TPR &
        0.934 & 0.938 & 0.945 \\ \hline Accuracy & 0.922 & 0.927 & 0.943 \\
    \hline \end{tabular} \label{allResults} \end{center} \end{table}

\begin{table}[htbp] \caption{SVM+KNN Ensemble Classifier Results} \begin{center}
    \begin{tabular}{|c|c|c|c|} \hline
        \textbf{Metric}&\multicolumn{3}{|c|}{\textbf{Sample Size}} \\
        \cline{2-4} & \textbf{6 Samples} & \textbf{12 Samples}& \textbf{24
        Samples} \\ \hline TNR & 0.804 & 0.836 & 0.926 \\ \hline FPR & 0.196 &
        0.164 & 0.074 \\ \hline FNR & 0.069 & 0.062 & 0.058 \\ \hline TPR &
        0.931 & 0.938 & 0.942 \\ \hline Accuracy & 0.921 & 0.930 & 0.940 \\
    \hline \end{tabular} \label{skResults} \end{center} \end{table}

\begin{table}[htbp] \caption{SVM+NB Ensemble Classifier Results} \begin{center}
    \begin{tabular}{|c|c|c|c|} \hline
        \textbf{Metric}&\multicolumn{3}{|c|}{\textbf{Sample Size}} \\
        \cline{2-4} & \textbf{6 Samples} & \textbf{12 Samples}& \textbf{24
        Samples} \\ \hline TNR & 0.939 & 0.953 & 1.000 \\ \hline FPR & 0.061 &
        0.047 & 0.000 \\ \hline FNR & 0.125 & 0.127 & 0.129 \\ \hline TPR &
        0.875 & 0.873 & 0.871 \\ \hline Accuracy & 0.880 & 0.879 & 0.881 \\
    \hline \end{tabular} \label{snResults} \end{center} \end{table}

\begin{table}[htbp] \caption{KNN+NB Ensemble Classifier Results} \begin{center}
    \begin{tabular}{|c|c|c|c|} \hline
        \textbf{Metric}&\multicolumn{3}{|c|}{\textbf{Sample Size}} \\
        \cline{2-4} & \textbf{6 Samples} & \textbf{12 Samples}& \textbf{24
        Samples} \\ \hline TNR & 0.982 & 0.993 & 1.000 \\ \hline FPR & 0.018 &
        0.007 & 0.000 \\ \hline FNR & 0.122 & 0.126 & 0.137 \\ \hline TPR &
        0.878 & 0.874 & 0.863 \\ \hline Accuracy & 0.886 & 0.883 & 0.873 \\
    \hline \end{tabular} \label{knResults} \end{center} \end{table}

Considering the results, the choice for the best approach is very dependent on
    the data set and desired outcomes.  The Naive Bayes and KNN classifiers are
    extremely fast, simple, and do well at maintaining low FPRs and moderate
    accuracy throughout the sample sizes.  Combining the SVM and KNN classifiers
    in an ensemble allowed the classifier to maintain greater than 90\%
    accuracy, but kept the FPR lower compared to using a SVM alone.  Thus, with
    very little data the best classification performance will come from a Naive
    Bayes or ensemble containing the Naive Bayes classifier such as the KNN and
    NB ensemble.  However, with sufficient data the SVM classifier alone still
    provides the best trade off between accuracy and FPR.

\section{Conclusion} \label{sec:conclusion} In this paper, we presented a
    quantitative comparison of four supervised machine learning algorithms'
    performance 
    when classifying ICs based on their ring oscillator network frequencies.  This method was able to achieve 97.6\%
    binary classification accuracy and a FPR of just 7.1\% when using a SVM
    classifier, and ensemble approaches achieved $\sim$88\% accuracy with no false
    positives. Despite these promising results, supervised learning approaches
    are often impractical in a real supply chain.  Finding proven 'Golden chips'
    is a challenge and knowing which chips are infected at the scaled assumed in
    the data set is near impossible.  Future work is planned to use unsupervised
    approaches and only "Golden" chip data to classify ICs as Trojan free or
    infected to maximize utility in real supply chains.
    \section{Acknowledgment} \label{sec:Acknowledgment} This work was supported
    in parts by the National Science Foundation under Grant Number CNS-1850241
    and UAH NFR. We would like to thank Dr. Tehranipoor and Dr. Forte for
    sharing the resources on Trusthub \cite{TRUSTHUB}.  \IEEEtriggeratref{14}
     \end{document}